\documentclass[prl,twocolumn]{revtex4}
\usepackage{graphicx}

\newcommand {\pmcube}      {m$^{-3}$}

\newcommand {\cmcube}      {cm$^3$}

\newcommand {\degrees}     {$^{\circ}$}
\newcommand {\mus}         {$\mu$s}

\begin{document}


\title{Dynamic and Stagnating Plasma Flow Leading to Magnetic Flux Tube Collimation}
\author{
   S. \surname{You},
   G. \surname{Yun} and
   P.~M. \surname{Bellan}
}
\affiliation{
   California Institute of Technology,
   Pasadena,
   California 91125,
   USA
}
\date{\today}

\begin{abstract}
   Highly collimated, plasma-filled magnetic flux tubes are frequently
   observed on galactic, stellar and laboratory scales. We propose that
   a single, universal magnetohydrodynamic pumping process explains why
   such collimated, plasma-filled magnetic flux tubes are ubiquitous.
   Experimental evidence from carefully diagnosed laboratory
   simulations of astrophysical jets confirms this assertion and is
   reported here. The magnetohydrodynamic process pumps plasma into a
   magnetic flux tube and the stagnation of the resulting flow causes
   this flux tube to become collimated.
\end{abstract}

\maketitle


The extreme collimation of astrophysical jets
\citep{meier01,deyoung91,burrows96} and the solar corona heating
mechanism \citep{walsh03} are two seemingly unrelated astrophysical
mysteries, yet both involve collimation of magnetic flux tubes.
Astrophysical observations \citep{deyoung91,burrows96} and
simulations \citep{meier01,nakamura01} indicate that bipolar plasma
outflows (jets) are natural \cite{meier01,lyndenbell03} features of
young stellar objects, black holes, active galactic nuclei and even
aspherical planetary nebula \citep{sahai03}. Although it has long
been presumed \citep{lovelace76,blandford82} that astrophysical jets
are magnetohydrodynamically driven, the standard models do not agree
on a single collimation process. A similar issue exists in solar
physics: solar spicules \citep{depontieu04}, prominences
\citep{solanki03,priest96} and coronal loops \citep{patsourakos04}
are considered to be plasma-filled filamentary magnetic flux tubes;
coronal heating models \citep{priest98,aschwanden01} then invoke
magnetic reconnection and plasma flow within such filamentary loops.
However, the models explain neither the origin of the observed flows
nor the extreme collimation (filamentary nature) of the observed
structures.

We propose that the collimation of any, initially flared,
current-carrying magnetic flux tube is due to the following process
\citep{bellan03}: a magnetohydrodynamic (MHD) force resulting from
the flared current profile drives axial plasma flows along the flux
tube; the flows convect frozen-in magnetic flux from strong magnetic
field regions to weak magnetic field regions; flow stagnation then
piles up this embedded magnetic flux, increasing the local magnetic
field and collimating the flux tube via the pinch effect. Thus, the
flux tube fills with ingested plasma and simultaneously becomes
collimated. This paper presents direct experimental evidence for
this process. We use ultra-high-speed imaging and Doppler
measurements of the fast plasma flows, combined with direct density
measurements before and after the filling of the flux tube.

\begin{figure}[t!]
  \includegraphics[width=0.75\columnwidth]{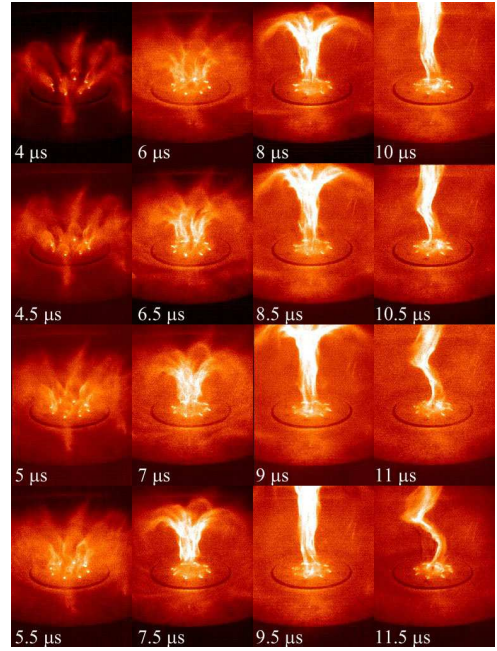}
  \caption{(log color) Typical plasma discharge
  sequence (\#6577, 2 million fps, 40~ns/frame).
  Eight collimated `spider legs' merge on axis to form
  a central column jet which collimates and propagates into the vacuum vessel.
  }
  \label{fig:shot sequence}
\end{figure}

Our experimental setup \citep{hsu02} simulates magnetically-driven
astrophysical jets at the laboratory scale by imposing boundary
conditions analogous to astrophysical jet boundary conditions (Fig.
\ref{fig:shot sequence}): a disk (cathode) representing a central
object such as a star, is coaxial and co-planar with an annulus
(anode) representing an accretion disk. A vacuum poloidal magnetic
field produced by an external coil links these two electrodes,
mimicking a poloidal magnetic field threading the accretion disk. A
radial electric field applied across the gap between the disk and
annulus drives poloidal current.

The nominal parameters of the lab experiment include a plasma
discharge duration $\sim20$~\mus, poloidal magnetic field
$B\sim0.01-0.3$~T, gun voltage $V_{gun}\sim1-7$~kV, poloidal current
$I\sim 50 - 200$~kA, density $\sim 10^{20-22}$~\pmcube. The
Alfv\'{e}n speed is $v_A\sim10^5$~m/s, and the ion gyroradius is
$\sim0.5$~mm, much smaller than the typical scale lengths ($L\sim1 -
50$~cm). The plasma $\beta$ ranges from $\ll$1 to $\sim$0.5,
assuming a plasma temperature $\sim10$~eV, and the Lundquist number
$S =\tau_r/\tau_A \sim 10^3 - 10^5$, depending on the characteristic
length used to define the resistive time  $\tau_r$ (estimated from
classical Spitzer resistivity) and the Alfv\'{e}n time $\tau_A$.
These dimensionless numbers are comparable to numerical MHD
simulations.
\begin{figure}
  \includegraphics[width=0.8\columnwidth]{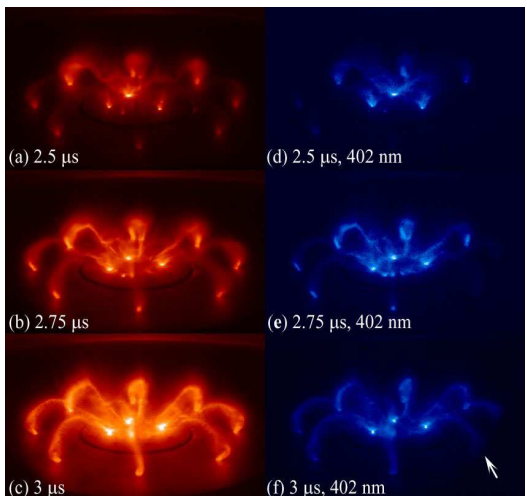}
  \caption{(color) Images from identical discharges. (a,d) 4345 (b,e) 4346
  (c,f) 4343. Nitrogen is puffed from the cathode nozzles and neon
  from the outer nozzles. Images (d-f) taken simultaneously with
  (a-c) resp., using a second camera and 402~nm filter
  to image nitrogen-rich plasma. Arrow indicates specific spider leg
  analyzed in detail in Fig.\ref{fig:plasma flow
  measurements}.
  }
  \label{fig:spider legs}
\end{figure}

\begin{figure}
  \includegraphics[width=0.84\columnwidth]{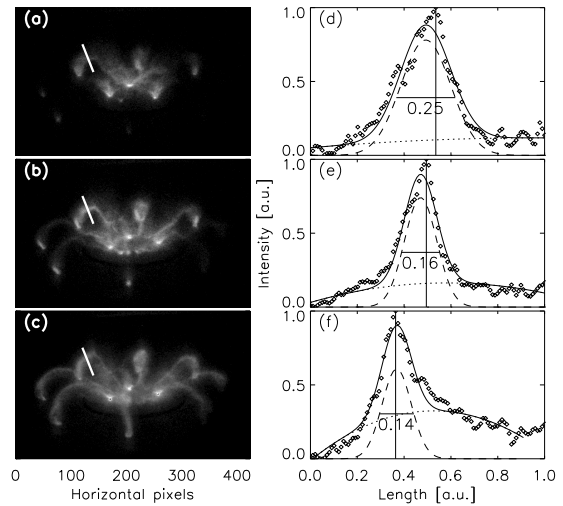}\\
  \caption{
    Spider-leg collimation measurements. (a-c) Images from Figs.
    \ref{fig:spider legs}d-f. Identical white lines indicate the
    measurements of the spider-leg intensity, panels
    (d-f). Dotted line shows the quadratic term of the least-squares
    fit (solid line), dashed line shows the Gaussian term. FWHM
    (number) is assumed to be representative of the spider-leg
    width.
  }
  \label{fig: spider leg collimation}
\end{figure}

Just before breakdown, neutral gas is puffed into the vessel from
nozzles located on the disk and the annulus. After breakdown, plasma
arches form, analogous to solar prominence loops \citep{bellan98}.
These arches are distributed toroidally, reminiscent of spider legs,
with each `leg' linking a gas nozzle on the disk to a corresponding
nozzle on the annulus (Fig.~\ref{fig:shot sequence}, 4.0~\mus~ and
Fig.~\ref{fig:spider legs}). These magnetically-dominated plasma
arches are initially flared (i.e. their minor radius increases with
distance along their axis). Because the coil-produced poloidal
vacuum magnetic field is weaker at the annulus than at the disk, the
outer footpoint diameter is initially 4-5 times that of the inner
footpoint. The eight legs quickly become bright and highly
collimated: in 0.5~\mus, as the legs fill up with plasma, the
outer/inner footpoint diameter ratio has been measured to reduce by
a factor of 2 (Fig. \ref{fig: spider leg collimation}), and can
approach unity to within $< 20\%$ (Fig. \ref{fig: flux tubes
flaring}). The legs later merge (Fig. \ref{fig:shot sequence},
5.5-7.5~\mus) to form a single central column (Fig. \ref{fig:shot
sequence}, 8.0~\mus), which constitutes the axially-expanding
`astrophysical jet' (Fig. \ref{fig:shot sequence}, 8-10~\mus). The
central column can eventually kink (Fig. \ref{fig:shot sequence},
10-11.5~\mus) and lead to spheromak formation \citep{hsu03}. The
initial flaring of the spider legs is just the flaring of the
coil-imposed vacuum magnetic field, a known analytic function
\cite{jackson_p178}.
\begin{figure}
  \includegraphics[width=0.84\columnwidth]{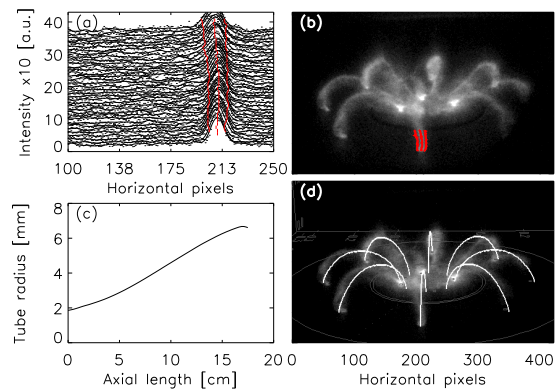}\\
  \caption{
    (color) (a, b) Measurement of flux tube flaring, on image \ref{fig:spider
    legs}f. FWHM (red marks) constant to within 20\%. (c) Analytical flaring of
    flux tube, computed from vacuum poloidal magnetic field
    \citep{jackson_p178}.
    (d) Comparison of the analytical flux tube axis positions (white lines) with image
\ref{fig:spider
    legs}f.
  }
  \label{fig: flux tubes flaring}
\end{figure}

\begin{figure}
  \includegraphics[width=0.6\columnwidth]{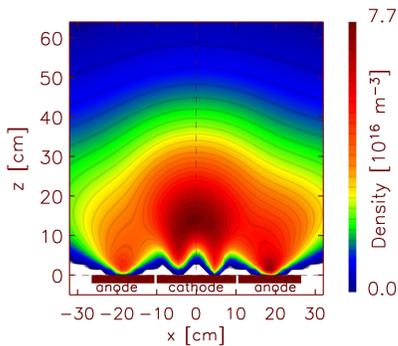}
  \caption{(color) Planar section of neutral density distribution just
  before breakdown, as measured by the fast ion gauge. Dashed line
  indicates the central axis of the experiment.
  }
  \label{fig:gas cloud section}
\end{figure}

\begin{figure}
  \includegraphics[width=0.84\columnwidth]{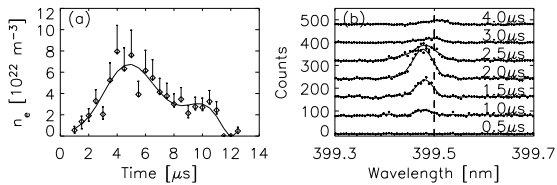}
  \caption{Spectroscopic
  measurements: (a) center column `astrophysical jet' $n_e(t)$
  inferred from Stark broadening. Solid line is a
  polynomial fit. Errors from 1-10 eV Doppler broadening and
  differing Lorentzian fit parameters \cite{mar00}. (b)
  Doppler shift of single spider-leg (N II, 399.5nm rest-frame, dashed
  line). Solid lines are polynomial fits. Times correspond to end
  of 1~\mus~exposure, relative to plasma breakdown.
  }
  \label{fig:spectroscopic measurements}
\end{figure}

Measurements shown in Fig. \ref{fig:gas cloud section} indicate that
the neutral gas density existing at breakdown is totally inadequate
to account for the plasma density measured in the spider legs or the
central column. The measurements were taken with a custom-built fast
ion gauge (FIG), with a 2\mus~response time, absolutely calibrated
with a standard commercial ion gauge. In situ measurements of the
gas cloud in front of the electrodes were fitted numerically to
obtain a complete 3D approximation of the gas output from the
nozzles. The numerical fit assumed a linear superposition of gas
cones from each nozzle, each with a pressure distribution having an
axial exponential drop and a radial Gaussian profile. In particular,
these measurements indicate that prior to breakdown, the neutral gas
density at the location of a single spider leg is
$\sim10^{17}$~\pmcube~(Fig. \ref{fig:gas cloud section}), which is
3-4 orders of magnitude less than the plasma density measured just a
few \mus~later (Fig. \ref{fig:spectroscopic measurements}a). These
time-resolved FIG measurements also show that, consistent with the
molecular thermal velocity of $\sim 1$~km/s, the characteristic
time-scale for neutral density evolution is $\sim10$~ms, so the
neutral gas distribution is quasi-stationary during the $<
50$~\mus~plasma discharge duration. For a plasma density $10^3-10^4$
times greater than the locally available supply of neutrals, simple
pinching of the tube would require $\sim$30-100 fold decrease in
radius. Only a 2-fold decrease is observed, so all spider leg
particles must be ingested from the gas valves into the leg by MHD
forces on \mus~time-scales.

\begin{figure}
  \includegraphics[width=0.84\columnwidth]{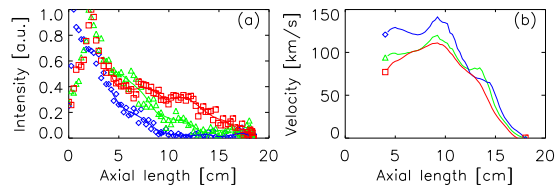}
  \caption{ (color)
    Plasma flow measurements: (a) Intensity along axis (of specific spider-leg shown in Fig. \ref{fig:spider legs}f) at 2.5~\mus~ (blue diamond, Fig. \ref{fig:spider legs}d),
    2.75~\mus~(green triangle, Fig. \ref{fig:spider legs}e), 3~\mus~(red squares, Fig. \ref{fig:spider legs}f). Solid lines are polynomial fits to the data between
    $4\le z\le 18$~cm. (b)  Plasma velocities at 2.625~\mus~(blue diamond line),
    2.75~\mus~(green triangle line) and 2.875~\mus~(red square line)
    obtained from Eq. \ref{eq: velocity from continuity equation}.
  }
  \label{fig:plasma flow measurements}
\end{figure}

To visualize this fast axial flow, nitrogen was injected from the
cathode footpoints of the spider legs and neon from the anode
footpoints (Fig. \ref{fig:spider legs}). A narrowband interference
filter (central wavelength 402~nm, Gaussian passband 10~nm) isolated
the nitrogen plasma for one camera (Fig. \ref{fig:spider legs}d-f),
while simultaneous, unfiltered photographs were obtained with a
second camera (Fig. \ref{fig:spider legs}a-c). A reconstruction of
the experimental geometry overlaid on the images allows comparison
of the dimensions of visible legs with best-fit vacuum magnetic flux
tube (Fig. \ref{fig: flux tubes flaring}d). The photographs indicate
a brightness propagating along the axis of the spider-leg flux tube,
from the cathode end to the anode end, on a 0.5~\mus~time-scale
(Figs. \ref{fig:spider legs} and \ref{fig:plasma flow
measurements}). This bright front propagation occurs
$>1.5$~\mus~after the breakdown process has ended, demonstrating
that it is different from streamer fronts \cite{zambra04} associated
with breakdown mechanisms. The time scale rules out ion acoustic
velocities ($8$~km/s) from consideration and the flow direction
rules out electrostatic acceleration. The propagation in fact
resembles a magnetized vacuum arc discharge \cite{boxman} and
involves nitrogen plasma flow, with $\sim10^2$~km/s velocities (Fig.
\ref{fig:plasma flow measurements}) and  a $\sim10^{11}$~m/s$^2$
mean acceleration. Flow velocities can be calculated assuming that
the density profile along the axis of the flux tube is proportional
to the square root of the light intensity, and integrating the
continuity equation
\begin{equation}
u(z)=\frac{1}{n(z)}\frac{\partial}{\partial t}\int_z^L
\frac{\partial n(z')}{\partial t}dz' \label{eq: velocity from
continuity equation}
\end{equation}
\noindent where the stagnation velocity $u(L) = 0$ and $L \sim
18$~cm is the spider leg's axial length (assumed to remain constant
during the measurements). The mean acceleration is estimated from
the typical velocity and time-scale ($u/t \sim 10^5/10^{-6} \sim
10^{11}$~m/s$^2$) or from the typical length-scale of the flow
($du_z/dt\sim
u^2/(2z)\sim10^{10}/(2\times5\times10^{-2})\sim10^{11}$~m/s$^2$).

According to the model \citep{bellan03}, the axial component of the
MHD pumping force is maximised on the flux tube axis and accelerates
plasma along this axis according to the equation of motion
\begin{equation}
\rho\left( \frac{\partial u_z}{\partial t}+u_z\frac{\partial
u_z}{\partial z} \right)=\frac{\mu_0 I^2}{2\pi^2 a^3}\frac{\partial
a}{\partial z} \label{eq: equation of motion}
\end{equation}
\noindent where $a$ is the local flux tube radius ($\partial
a/\partial z$ is the flaring), $I$ is the total current in the flux
tube, $u_z$ is the axial plasma velocity and $\rho$ is the plasma
density.

Using spider-leg flux tube dimensions measured from the photographs
(Fig. \ref{fig: flux tubes flaring}c, $a\sim4$~mm, $\partial
a/\partial z\sim0.02$) and assuming the current inside a single leg
is one eighth the total measured 120~kA gun current, Eq.\ref{eq:
equation of motion} predicts a spider-leg plasma density of
$\sim2\times10^{21}$~\pmcube. This implies that within 0.5~\mus, MHD
forces inject about $2\times10^{16}$~nitrogen particles from the gas
nozzles into a $\sim10$~\cmcube~spider-leg. Summing over all eight
flux tubes gives a total influx pumping rate of
$\sim3\times10^{17}$~particles/\mus. Since it takes
$\sim10$~\mus~for the legs to merge and form the 2.5~cm diameter,
50~cm long central column, assuming the pumping rate is constant,
the central column should then have a
$\sim1\times10^{22}$~\pmcube~volume-averaged plasma density. The
central column jet, located where 10~\mus~previously there was
essentially vacuum, has ingested about 10\% of the total number of
particles available from the gas feed lines. At this rate, it would
take a few hundred  $\mu$s to empty all the gas from the feed lines.
This time is much shorter than the time-scale of fusion experiments,
where similar filamentary flux tubes have been observed to emerge
\citep{wilson04} explosively from the hot plasma core and connect to
the cold walls.
\begin{figure}
  \includegraphics[width=0.84\columnwidth]{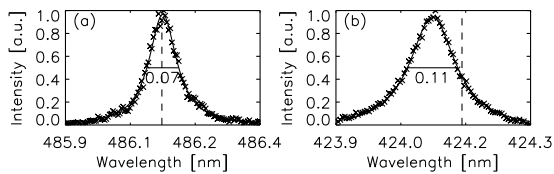}
  \caption{Measured line spectra with Lorentzian fits: (a) single spider-leg, H
  486.133 nm (dashed line). Exposure 0-1~\mus~w.r.t breakdown. Line of
  sight is perpendicular to the spider-leg, parallel to the cathode
  plane.
  (b) Fully formed central column jet (N II, dashed line is 424.178 nm rest-frame wavelength).
  }
  \label{fig:line spectrum measurements}
\end{figure}

The prediction for very high plasma densities in the spider legs and
the central column has been substantiated using independent
spectroscopic line measurements (Figs. \ref{fig:spectroscopic
measurements}a, \ref{fig:line spectrum measurements}). These
measurements can exhibit large Stark broadening indicative of high
electron densities. For hydrogen plasmas, we measure
$\sim2\times10^{20}$~\pmcube~in the spider-leg (Fig. \ref{fig:line
spectrum measurements}a) and $\sim2\times10^{22}$~\pmcube~in the
fully-formed central column jet. In nitrogen discharges, central
column jet densities $8\times10^{22}$~\pmcube~are measured (Fig.
\ref{fig:spectroscopic measurements}a at 4~\mus, Fig. \ref{fig:line
spectrum measurements}b) however spider-leg densities cannot be
resolved. Spectral lines were sampled by optical fibres viewing
selected portions of the discharge and recorded with an optical
multichannel analyzer. The FWHM $g$ [nm] is related to $n_e$
[\pmcube] by $g = 1.6\times10^{-24} n_e$ for NII \citep{mar00} and
$g = 2\times10^{-15} n_e^{2/3}$ for H \citep{griem}. The
measurements are corrected for the instrument function and a 1-10~eV
thermal broadening, estimated from ion line ratios.

The pumping model also predicts that the largest plasma flow
velocity in the tube is of the order of the Alfv\'{e}n velocity
associated with the magnetic field generated by the axial current.
Here, the 15~kA current, $4\times10^{20-21}$~\pmcube~density and
4~mm typical radius of the spider-leg gives $v_A\sim70-220$~km/s,
encompassing the measured $\sim100$~km/s shown in Fig.
\ref{fig:plasma flow measurements}. On the axis of a single
spider-leg (Fig. \ref{fig:spectroscopic measurements}b), we observe
blue shifts corresponding to $\sim20-40$~km/s line-of-sight
velocities. Using CCD images, we estimate the optical probe to be
8~cm along and 1~cm away from the spider-leg axis, with a
line-of-sight angle of 72\degrees$\pm$3\degrees to the plasma flow
direction. In the central column jet, the flow velocity is measured
at $\sim40-70$~km/s (Fig. \ref{fig:line spectrum measurements}b),
with a line-of-sight coincident to the jet axis. These velocities
are consistent with the Alfv\'{e}n velocity expected from the
measured total 150~kA current and $7\times10^{22}$~\pmcube~peak
density in the central column. Both the collimation of the spider
legs and the central column jet arise from the same process.

The results thus provide strong evidence for flow-driven collimation
of flared magnetic flux tubes. This process requires net electrical
current, a supply of particles and stagnation of the driven flow. In
an astrophysical context, accretion disks supply plasma and flows
stagnate at the lobes; a net axial current is understood to be
necessary for self-confinement \citep{ferrari98} and recent
observations of helical magnetic field structure in jets are
indicative of net axial current \citep{uchida04}. The electrical
circuit is then closed by radial currents in the accretion disk,
returning via the cocoon. Active solar regions also exhibit
unneutralized electrical currents \citep{wheatland00}. We emphasize
that strong flows and collimation can be driven by even modest
amounts of current, before reaching the kink instability (high
current) threshold. Supported by the US DoE and NSF.



%


\bibliography{../bibliography/books,../bibliography/articles}

\end{document}